# Evaluation of Delay Uncertainty in PCB Interconnects Due to Fiber Weave Effect

Alex Manukovsky[1], Yuriy Shlepnev[2], Shimon Mordooch[3]

[1]*Inte,l alex.manukovsky@intel.com;* [2]*Simberian Inc., shlepnev@simberian.com;*
[3]*Majore Tech Company, shimon.mordooch@gmail.com*

*Abstract* — **Delay Deviation Exceedance (DDE) and Differential Skew Exceedance (DSE) measures are proposed to quantify delay uncertainty in single-ended and differential PCB interconnects arising from Fiber Weave Effect (FWE). DDE and DSE are constructed with 3D EM analysis of traces over inhomogeneous dielectric with glass fiber bundles in resin. Measurements or 3D EM models for FWE are usually used for observations of delay or impedance dependency on position of traces over fiber bundles with the purpose to find the worst case scenario. This paper turns the results of 3D EM analysis into probabilistic measures of possible delay or skew uncertainty – probability to have delay deviation in single-ended interconnects over allowed limit (DDE measure) or to have skew over allowed limit in differential interconnects (DSE measure). The introduced measures allow formalizing the laminate selection process for parallel as well as for serial PCB interconnects.**

## I. INTRODUCTION

Low-cost Printed Circuit Board (PCB) laminates are usually manufactured with glass fabric and resin with different dielectric properties. It creates inhomogeneous dielectric with lattice type of variations in dielectric permittivity. This inhomogeneity introduces variations and uncertainties in interconnects delay, skew and impedance. The uncertainty in flight time in single-ended interconnects of parallel buses such as DDR and skew uncertainty in differential serial buses such as Ethernet may cause unpredictable signal degradation and data link failures. This unwanted variation in electrical parameters of interconnects is called Fiber Weave Effect (FWE). FWE was a subject of intense investigation since early 2000s [1]-[6]. Over that time, the laminate industry came up with more homogeneous PCB materials almost free from the FWE. Also different interconnect routing techniques were suggested to mitigate the FWE. However, all that increases the cost of PCB manufacturing that is also unwanted factor, especially for a large volume electronics manufacturing.

The increasing data rates both in parallel and serial buses put additional constrains on the unwanted uncertainty in the delay or skew. Low-cost PCB materials can exhibit up to 10 ps/inch of delay or skew variations [4]-[6]. How it can affect interconnect performance if no FWE mitigation technique is used? One of the key properties of a successful DDR bus design is maintaining the same flight time across all data lanes within the signal group. For DDR5 technology operating in 6400 MT/s this number would be typically around 20-50 ps. For DDR6, maximum data rate peaks above 12800 MT/s, hence the bound is expected to be a half of that. Meeting this flight time constraint within data signals poses several practical challenges. One of them, overlooked in the previous generations of DDR, is delay uncertainty induced by glass weave in PCBs. Serial links are usually implemented as the differential interconnects and the latest serial data transfer standards have single links operating over 56 Gbps with the single bit duration of about 17.85 ps for NRZ and 35.71 ps for PAM4 signaling. Considering the length of such links even a few ps per inch of a skew uncertainty would be unacceptable.

The goal of this paper is to quantify delay uncertainty in single-ended links and skew uncertainty in serial differential links and introduce a formal measure which can be used in selection of PCB materials. In this paper, a new Delay Deviation Exceedance (DDE) measure is proposed to quantify the delay uncertainty in single-ended links and Differential Skew Exceedance (DSE) is proposed to quantify the uncertainty in differential links. DDE and DSE are constructed with 3D EM analysis of traces over an inhomogeneous dielectric. 3D EM models were often used to simulate the FWE - see [4], [5] and references there. However, such models are used to simulate either particular cases, or simulate the delay or impedance dependence on position of traces over fiber bundles and find worse-case scenarios. The question is what is the probability to have some number of worst cases in a particular case? The answer is provided through DDE and DSE measures proposed in this paper. Note that the authors of [3] were on the way of constructing such uncertainty measure – they constructed probabilistic model of skew that was not normal and used it to find probability to have the skew over some limits. However, the process was not formalized for the practical applications as it is done here.

## II. 3D EM MODEL OF FWE

PCB laminates are usually made of woven fabric impregnated with resin. The low-cost fabrics use glass fiber bundles and may have different styles of weave. Dielectric properties of the glass bundles and resin may be considerably different and the precise analysis requires simulation of complicated geometries of the fabric material in the resin dielectric [4]-[5]. Though, the problem can be simplified without loss of the essential important effects of periodic dielectric inhomogeneity. Instead of reproducing particular fabric style geometry, we use simplified 3D geometry as shown in Fig. 1. The original dimensions of the glass bundles are X1-X3 and Y1-Y3 taken from measurements provided in [6] and rounded off to values X1'-X3', Y1'-Y3' shown in Table 1, to fit Cartesian mesh. Model Parameters X3' and Y3' define the period of the dielectric lattice. Parameters X2', Y2' are size of the bundles along the X and Y axes and are further adjusted in the model to have about the same volume of glass in resin (elliptic shape of each bundle is transformed into rectangular shape to have the same area). Parameters X1'=Y1' define size of the bundles along the Z-axis. Dielectric constant for the glass is set to 6 and for resin is set to 3.5. The resulting structure has same features and the original geometry – areas of overlapping bundles ("glass hills" - double size along the Z-axis), areas with only resin ("resin valleys") and areas with just X-directed or Y-directed bundles. Laminate thickness H is set to 4 mil and trace thickness T is set to 0.75 mil in all examples.

blue stars and the interpolated values by brown lines. Probability density is computed with 100000 samples and 20 bins.

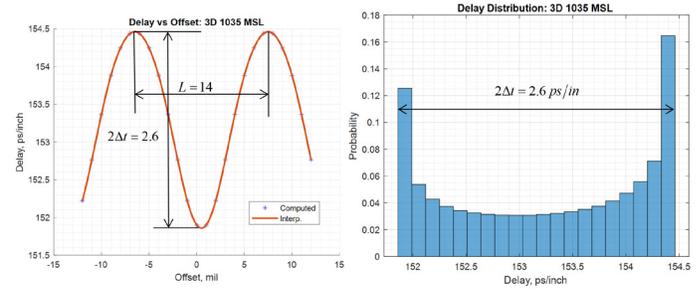

*Fig. 2. Delay variation with offset (left plot) and corresponding delay probability density histogram (right plot) for 1035 fabric style.*

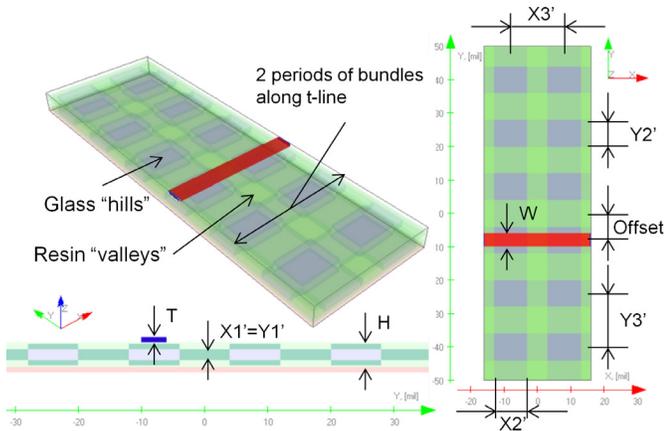

*Fig. 1. 3D EM model of single-ended microstrip segment over dielectric composed of glass and resin.*

Table 1. Model fiber weave bundle parameters (all in mils).

| Style | X1' | X2' | X3' | Y1' | Y2' | Y3' |
|-------|-----|-----|-----|-----|-----|-----|
| 1035  | 0.8 | 9   | 14  | 0.8 | 12  | 14  |
| 1080  | 1.35| 8   | 17  | 1.35| 12  | 22  |
| 1078  | 1.2 | 14  | 16  | 1.2 | 17  | 18  |
| 3313  | 1.7 | 13  | 16  | 1.7 | 11  | 16  |

Trefftz Finite Elements (TFE) method is used to build 3D EM models of the problem [7]. The method is implemented in Simbeor software [8] as 3DTF solver. Single-ended and differential trace segments are simulated at 10 GHz with the offset parameter swept to extract the phase delay and impedance variations over at least one period of the glass lattice. To have phase delay independent of the reflections, it is extracted from S-parameters normalized to characteristic impedance of the periodic structure (reflection-less). All computations are automated in FEW Kit provided with Simbeor SDK software [8].

### III. DELAY VARIATION AND PROBABILITY

Four laminate cases with the parameters shown in Table 1 are simulated for with trace width W=4 mil and offset ranging from -12 to +12 mils and offset step 1 mil. The phase delay per unit length is computed and interpolated with cubic splines. Delay probability density is then evaluated assuming the uniform distribution of the trace offsets. The results are shown in Fig. 2 – Fig. 5. Computed delay values are shown by

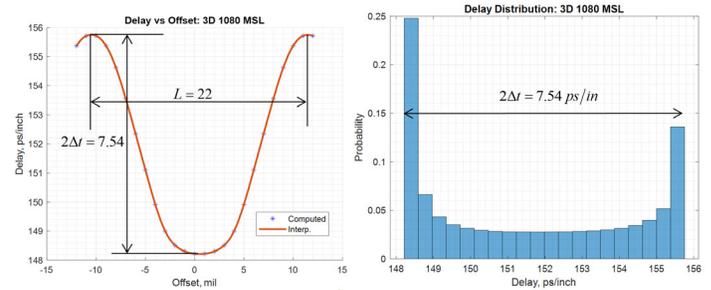

*Fig. 3. Delay variation with offset (left plot) and corresponding delay probability density histogram (right plot) for 1080 fabric style.*

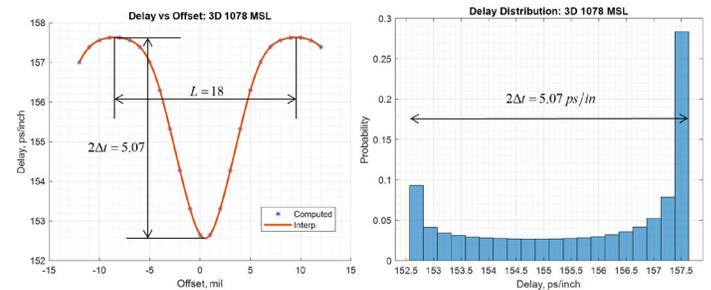

*Fig. 4. Delay variation with offset (left plot) and corresponding delay probability density histogram (right plot) for 1078 fabric style.*

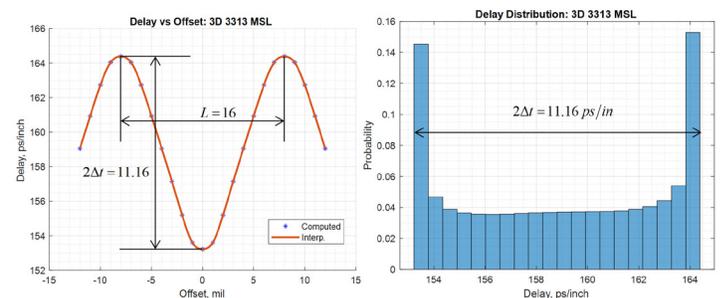

*Fig. 5. Delay variation with offset (left plot) and corresponding delay probability density histogram (right plot) for 3313 fabric style.*

The delay variation is a periodic function with the period $L$ equal to the period of the original lattice model Y3'. The corresponding delay probability is bounded by the minimal and maximal values (worst cases). The probability to have the minimal and maximal delay values is the highest in all 4 cases. Notice that cases 1035 and 3313 have delay variation close to sinusoidal and corresponding probability density is similar to the Arcsine distribution. Fabric style 1080 has larger "resin valley" areas and corresponding delay dependency has flatter bottoms and higher probability to have smaller delay. On the other hand, fabric style 1078 has larger "glass hill" areas and higher probability to have larger delay.

Measured variations in effective dielectric constant close to sinusoidal are observed in measured data in [1]. Though the corresponding probability densities for the effective dielectric constant in [1] were not close to the Arcsine due to either additional random variations in the measurements (should be accounted with the kernel density estimate) or not uniform distribution of the trace positions (there was some order in arranging trace segments on the PCB). Dielectric constant variations extracted from measured data in [6] are close to sinusoidal or clipped sinusoidal.

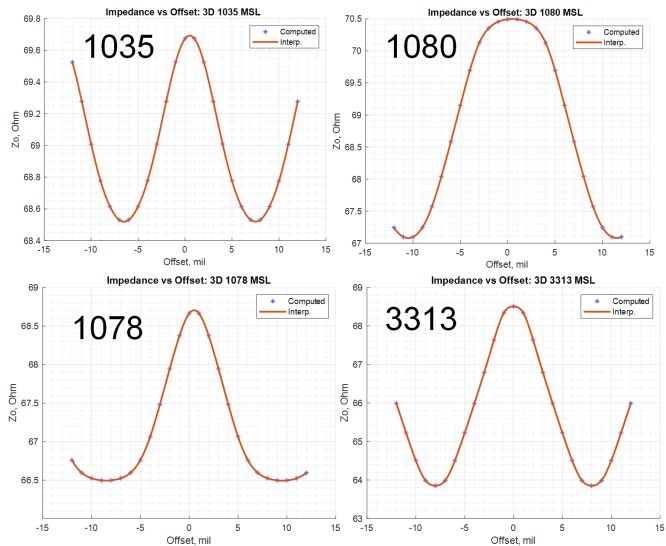

*Fig. 6. Impedance variation with the offset for 4 mil trace for four fabric styles. Computed values are shown be stars and interpolated by line.*

IV. DELAY DEVIATION PROBABILITY AND EXCEEDANCE

The delay variations with the offset and corresponding probability densities are computed and presented in the previous section. It is clear that the probability to have cases close to the worst case scenario may be quite high. Those are useful results, but the goal is to have a quantity to characterize the uncertainty in the delay due to FWE. Complimentary Cumulative Distribution Function (CCDF) computed for the delay deviation probability density can be used as such measure of uncertainty. Technically, it is probability to have delay deviation larger than certain specified value. It is called Delay Deviation Exceedance or DDE. The delay deviation probability density is computed first and shown on the left plots in Fig. 7-10. Then, DDE values are computed for values N ps/inch and shown on the right graphs in Fig. 7-10.

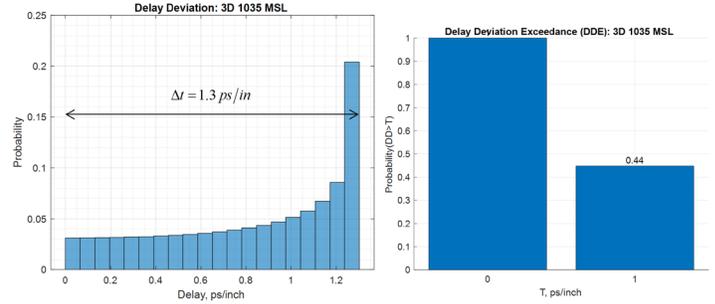

*Fig. 7. Delay deviation probability density (left plot) and corresponding DDE (right plot) for 1035 fabric style.*

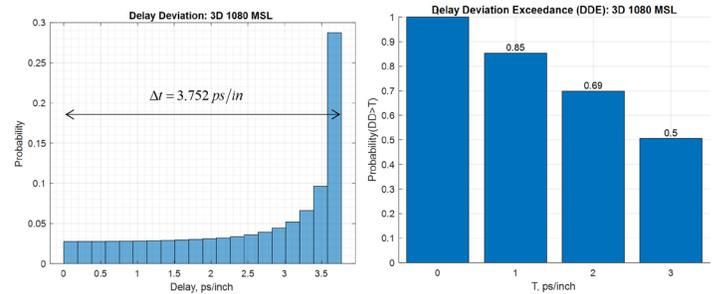

*Fig. 8. Delay deviation probability density (left plot) and corresponding DDE (right plot) for 1080 fabric style.*

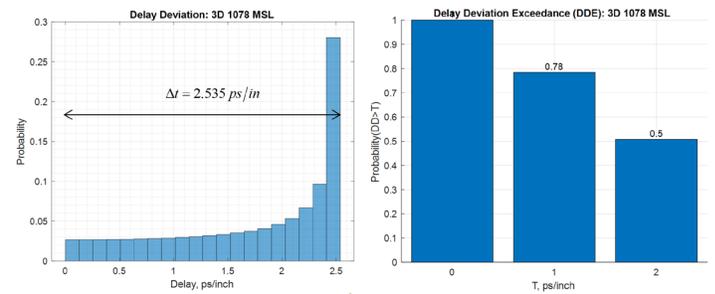

*Fig. 9. Delay deviation probability density (left plot) and corresponding DDE (right plot) for 1078 fabric style.*

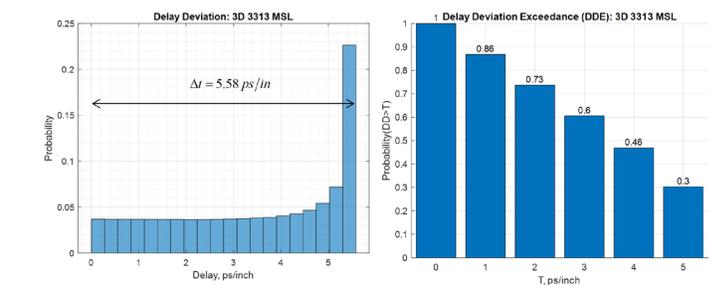

*Fig. 10. Delay deviation probability density (left plot) and corresponding DDE (right plot) for 3313 fabric style.*

Comparison of DDEs for all four fabric styles is show in Fig. 11. Now we can see if DDR specification does not allow the delay uncertainty over 3 ps/inch for instance, fabric styles 1080 and 3313 cannot be used without some FWE mitigation technique (routing at angle or panel rotation). The expected number of cases with the delay deviation over 3 ps/inch is about 50% for 1080 fabric and 60% for 3313 fabric.

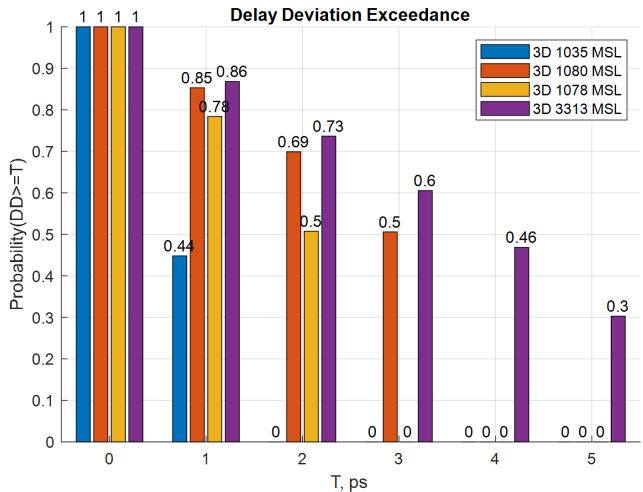

Fig. 11. Comparison of DDEs for four fabric styles.

## V. DIFFERENTIAL SKEW VARIATION, PROBABILITY AND EXCEEDANCE

Four laminate cases with the parameters shown in Table 1 are simulated with two parallel traces with equal width W=4 mil separated by S=4 mil and offset ranging from -12 to +12 mils and the offset step 1 mil (geometry is similar to shown in Fig. 1). The offset coordinate is exactly between two traces. The skew is defined by the difference of phase delays computed for each trace. Computed dependencies of the skew values from the offset coordinate are shown in Fig. 12 by blue stars and the interpolated values by brown lines for all 4 laminate cases. Similar skew variations with the offset close to sinusoidal were also observed in [3]-[5].

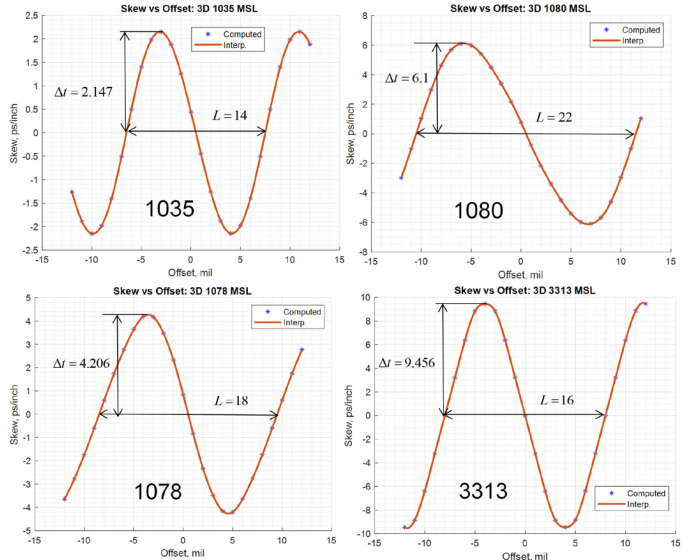

Fig. 12. Computed differential skew variations with the offset for 4 laminates.

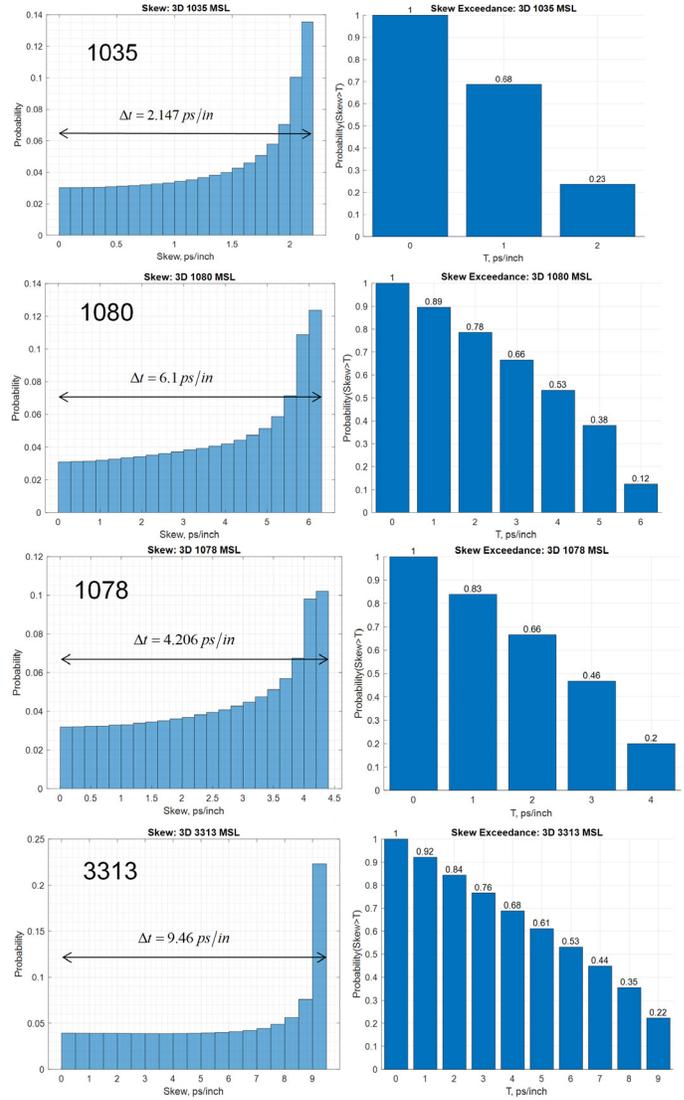

Fig. 13. Differential skew probability density (left) and DSEs computed for 4 laminates.

With the skew dependency on the offset, skew probability density is then evaluated assuming the uniform distribution of the traces offsets. Probability density is computed with 100000 samples and 20 bins and is shown in Fig. 13. The fact that the skew probability density due to FWE is not normal is also noticed in [3].

Finally, comparison of DSEs for all four fabric styles is show in Fig. 14. Now we can see if DDR specification does not allow the skew uncertainty over 3 ps/inch for instance, the only fabric style 1035 can be used without additional skew mitigation techniques. The expected number of cases with the delay deviation over 3 ps/inch is about 44% for 1078 fabric and 66% for 1080 and 76% for 3313 fabrics – that is not negligible by any means.

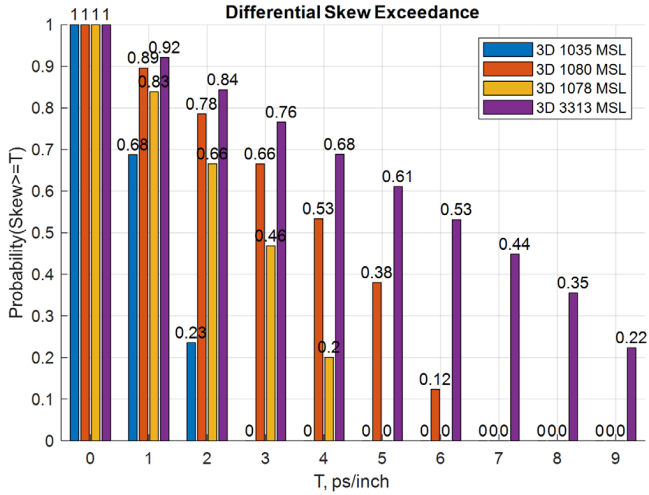

*Fig. 14. Comparison of DSEs for four laminates.*

## VI. ANALYTICAL MODEL

Possible delay deviation or skew approximation is the sine function:

$$DD(x) = \Delta t \left| \sin\left(\frac{2\pi x}{L} + \alpha\right) \right|, \quad x \in [-L/4, +L/4]$$

Here $L$ is the period and $\Delta t$ is the amplitude or maximal possible deviation of the delay or skew.
Corresponding probability density function is as follows:

$$P(t) = \frac{2}{\pi \cdot \Delta t \sqrt{1 - \left(\frac{t}{\Delta t}\right)^2}}, \quad t \in [0, +\Delta t]$$

It has Complimentary CDF defined by the arcsine function (arcsine distribution [9]):

$$S(t) = P(T \geq t) = 1 - \frac{2}{\pi}\arcsin\left(\frac{t}{\Delta t}\right), \quad t \in [0, +\Delta t]$$

It is the probability to have delay deviation or skew over certain limit. It can be used for approximate evaluation of the DDE or DSE and requires just one parameter identification – the maximal possible deviation $\Delta t$ (worst case). Comparison of the DDEs and DSEs computed directly from numerical experiment and from the arcsine distribution is provided in Fig. 15 for DDEs and in Fig. 16 for DSEs. Instead of the arcsine distribution, Beta or Kumaraswamy [10] distribution can be used for better accuracy. However, it will require identification of two or more parameters, instead of one in arcsine.

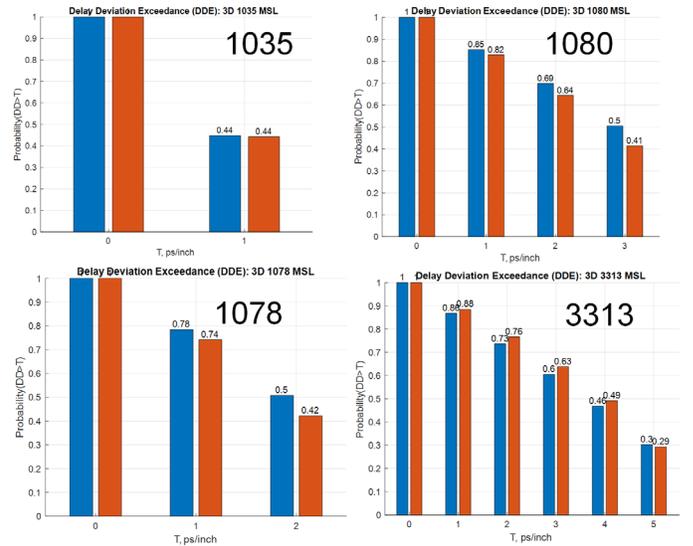

*Fig. 15. DDEs computed directly from numerical experiment for single trace (blue bars) and from the arcsine CCDF approximation (brown bars).*

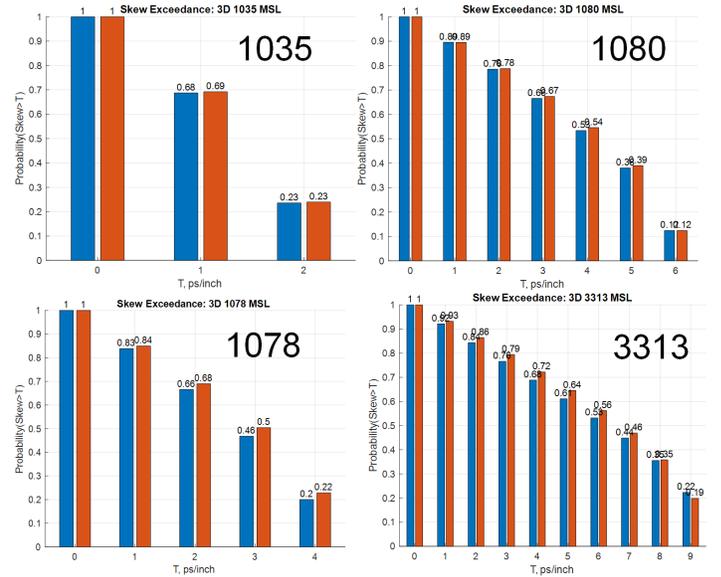

*Fig. 16. DSEs computed directly from numerical experiment with differential traces (blue bars) and from the arcsine CCDF approximation (brown bars).*

## VII. CONCLUSION

A new Delay Deviation Exceedance (DDE) measure is proposed to quantify the delay uncertainty in single-ended links and similar Differential Skew Exceedance (DSE) measure is proposed to quantify the uncertainty in differential links. DDE and DSE are computed with numerical experiment by running multiple 3D EM simulations of short segments of interconnects over inhomogeneous dielectric. Delay deviation and skew probability densities are evaluated from interpolated dependencies of the delay deviation and skew from offset of trace or traces with respect to the glass bundles. The exceedances are then computed as the complimentary CDFs from the corresponding probability densities. Note that the results of such analysis will depend on the trace width and separation in addition to the geometry of the laminate itself. Trace averages the dielectric properties. Wider traces will see less variations comparing to narrow traces. Thus, a numerical experiment or measurements should be done for each practical case. All computations done in this paper are easily repeatable with FWE Kit of Simbeor SDK [8].

It is shown that the arcsine distribution can be used for approximate preliminary evaluation of the DDE and DSE. It requires only the worst case delay deviation or worst case differential skew. Those parameters can be obtained from just two numerical experiments or just two measurements – for trace over the "glass hills" and trace over the "resin valleys".